\documentclass[12pt]{iopart}
\usepackage{graphicx}
\usepackage{color}
\usepackage[utf8]{inputenc}
\usepackage{subcaption}
\expandafter\let\csname equation*\endcsname\relax

\expandafter\let\csname endequation*\endcsname\relax

\usepackage{amsmath}
\usepackage{graphicx} 
\usepackage{iopams} 
\usepackage{url}
\newcommand{\commentout}[1]{}

\begin{document}
\title[Neutron Capture Rates]{Capture Rates of Highly Degenerate Neutrons}

\author{B. Knight$^{1, *}$, O. L. Caballero $^1$, H. Schatz$^{2,3,4}$}

\address{$^1$ Department of Physics, University of Guelph, Guelph, ON, Canada, N1G 2W1}
\address{$^2$ National Superconducting Cyclotron Laboratory, Michigan State University, East Lansing, MI 48824 USA}
\address{$^3$ Joint Institute for Nuclear Astrophysics- Center for the Evolution of the Elements (JINA-CEE), USA}
\address{$^4$ Michigan State University, East Lansing, MI 48824 USA}
\address{$^*$ Author to whom any correspondence should be addressed}
\ead{knightb@uoguelph.ca,ocaballe@uoguelph.ca,schatz@frib.msu.edu}
\vspace{10pt}
%\begin{indented}
%\item[]August 2017
%\end{indented}

\begin{abstract}
At the low temperature and high density conditions of a neutron star crust neutrons are degenerate. In this work, we study the effect of this degeneracy on the capture rates of neutrons on neutron rich nuclei in accreted crusts. We use a statistical Hauser-Feshbach model to calculate neutron capture rates and find that neutron degeneracy can increase rates significantly. Changes increase from a factor of a few to many orders of magnitude near the neutron drip line. We also quantify uncertainties due to model inputs for masses, $\gamma$-strength functions, and level densities. We find that uncertainties increase dramatically away from stability and that degeneracy tends to increase these uncertainties further, except for cases near the neutron drip line where degeneracy leads to more robustness. As in the case of capture of classically distributed neutrons, variations in the mass model have the strongest impact. Corresponding variations in the reaction rates can be as high as 3 to 4 orders of magnitude, and be more than 5 times larger than under classical conditions. To ease the incorporation of neutron degeneracy in nucleosynthesis networks, we provide tabulated results of capture rates as well as analytical expressions as function of temperature and neutron chemical potential, for proton numbers between $3 \le Z \le 85$, derived from fits to our numerical results. Fits are based on a new parametrization that complements previously employed power law approximations with additional Lorentzian terms that account for low energy resonances, significantly improving accuracy. 

\end{abstract}


\submitto{\jpg}

\section{Introduction}
Neutron stars in binary systems can accrete light elements such as hydrogen and helium from the companion star into their envelope. The accreted material undergoes hydrogen and helium burning, for example via the rp-process, producing new elements that change the original composition of the neutron star crust \cite{Wallace:1981zz,Schatz:1998zz,Schatz:1999kx,Meisel2018}. The interpretation of astronomical observations from these binary systems, such as X-ray bursts, superbursts \cite{Strohmayer, Liu:2007ts,Galloway:2017jcn}, transiently cooling neutron stars \cite{Meisel2018}, and potentially gravitational waves \cite{Patruno:2017oum}, rely on our understanding of the composition and nuclear reactions occurring in the accreted crust.

Once the rp-process comes to an end, the ashes are pushed deeper into the star by the ongoing accretion setting the initial composition for further reactions \cite{1990A&A...227..431H,Haensel:2003ti,Gupta:2006fd}. Neutron captures become a prominent reaction mechanism when the ashes are buried in the sea of highly degenerate neutrons present in the inner crust. Prior to the neutron drip line, electron capture induced neutron emission in the outer crust can already lead to the appearance of free neutrons that can be re-captured \cite{Lau:2018put}. These processes alter the composition and subsequent reactions in deeper layers. 

While large scale neutron capture rate compilations \cite{Holmes:1976bba,Rauscher:2000fx} based on Maxwellian averaged cross sections applicable to stellar helium burning and typical $r$-process sites are readily available \cite{Pritychenko2020, Dillmann:2008fs, Cyburt:2010, Xu:2012uw}, there is a scarcity in studies relevant to accreting neutron stars where neutron degeneracy has a stronger impact.  Neutron degeneracy has been demonstrated to have a large effect on capture rates for Mg and Ca isotopes, together with plasma effects due to the degenerate electron gas \cite{Shternin:2012pt}. These authors developed an analytical approach for degenerate capture rates assuming that the cross sections can be approximated by a power law fit, which allows to calculate corrections to Maxwell-Boltzmann averaged rates in a straight forward manner. The fitting to a power law is motivated by the $1/\sqrt{E}$ trend of the low energy capture cross section. The authors also found, for the reactions studied, that endothermic rates were orders of magnitude larger than their classical counterparts. Furthermore, as was shown in \cite{Knight:2019UNI} and is further discussed later in this paper (section \ref{fit}), with a lack of experimental information, a power law approach can result in large uncertainties related to the "goodness" of the fit, compounded further by uncertainties in the nuclear input. Here we calculate neutron capture rates under degenerate neutron conditions by integrating neutron energy distributions over cross sections calculated with the Hauser-Feshbach statistical model  \cite{Hauser:1952zz} TALYS \cite{talys} for the broad range of nuclei relevant for neutron star crust processes. We also provide an analytical expression for the cross sections that builds on the power law fit of \cite{Shternin:2012pt}, by including possible low-energy resonances that are potentially sampled at high chemical potentials; thus facilitating the implementation of our results in nucleosynthesis studies wherever the neutron degeneracy is high. In this regard, degenerate capture rates were incorporated by Lau et al \cite{Lau:2018put} using the methodology of Sthernin et al \cite{Shternin:2012pt}, in their extensive study of nuclear reactions in accreted crusts. The results presented here show significant differences compared to the pioneer work of Ref. \cite{Shternin:2012pt} motivating analogous work to that of Ref. \cite{Lau:2018put}.

An important need for astrophysical applications are estimates of the uncertainties of the neutron capture rates employed, in particular for the reaction rates on nuclei far from stability where no experimental data exist. To that end,    
 several studies have addressed the effect of different nuclear physics inputs on Maxwell-Boltzmann averaged neutron capture rates for explosive nucleosynthesis as well as the \textit{i}-process \cite{Bertolli:2013gka,Denissenkov:2018gph,Liddick:2016eot}. An important finding was the dramatic increase of uncertainties for isotopes farther from stability. The impact of these changes on the r-process was studied in e.g. \cite{Liddick:2016eot, Arcones:2010dz,Mumpower:2012qy,Surman:2013wfa}. In the context of uncertainties it is also worth mentioning that different code implementations of the statistical models result in different cross sections, when the nuclear physics input was the same, as shown by Beard et al \cite{Beard:2014jaa}. In this paper we determine the impact of nuclear physics input variations on capture rates of degenerate neutrons. 

This paper is organized as follows: in section \ref{sec:neutron captures} we introduce neutron degeneracy in capture rates, study the variability of such rates to nuclear physics input and to different degenerate conditions. In section \ref{fit} we present our analytical expression for the cross section and rates, and quantify the uncertainty introduced by this approach. Discussion and concluding remarks follow in section \ref{sec:discussion}.

%===================================================================

\section{Neutron capture rate calculations}
\label{sec:neutron captures}
Thermonuclear neutron capture rates are calculated by statistically averaging the capture cross sections $\sigma^*$ of a reaction $X+n\rightarrow Y+\gamma$ over the relative velocity ($v$) distribution function $f(v)$ of the neutron-target system. We assume a non-relativistic collision energy  $E=m v^2/2$, where $m$ is the reduced mass of the system, which for most of the reactions of interest here, is very close to the neutron mass $m\approx m_n$, and so $E$ is the energy of the neutron relative to the target. Thus we can write the reaction rate as
\begin{equation}
\langle \sigma^*v\rangle=\sqrt{\frac{2}{m}}\frac{1}{N}\int_0^\infty E \sigma^*(E) f(E)dE,
\label{averagesigma}
\end{equation}
where $f(E)$ is the neutron distribution function and  $N$ is the normalization factor given by
\begin{equation}
N=\int_0^\infty \sqrt{E} f(E)dE.
\end{equation}

Depending on the system energy scales and thermodynamic conditions the distribution function can take the form of the classical Maxwell-Boltzmann distribution, $f(E)_{MB} = e^{-E/T}$ where $T$ is the temperature in MeV, or to account for degeneracy, a Fermi-Dirac distribution $f(E)_{FD}={[1+e^{(E-\mu_{n})/T}]}^{-1}$ where $\mu_{n}$ is the neutron chemical potential. In general, the total cross section, $\sigma^*$, can include contributions from thermally excited states of the target nuclei $X$,

\begin{equation}
\sigma^*(E)=\frac{\sum_a g_a \exp(-E^{(a)}_X/T) \sum_b \sigma_{ab}(E)}{\sum_a \exp(-E^{(a)}_X/T)},
\end{equation}  
where $E^{(a)}_X$ is the energy of the level $a$, $g_a=(2J+1)$ is its statistical weight based on its spin $J$, and $\sigma_{ab}$ is the partial cross section for the reaction $X^{(a)}+n \rightarrow  Y^{(b)} + \gamma$, with $a$ and $b$ denoting energy levels of the target and residual nuclei respectively. However, given the low temperature of the crust (typically $T<40$~keV) compared to typical excitation energies of low lying states
we ignore the contribution from excited states in the sum above, and take  $\sigma^* = \sigma$, where $\sigma$ is the ground state cross section, as proposed in Ref. \cite{Shternin:2012pt}. Degeneracy effects quickly taper to the classical results with increasing $T$, justifying this approach. In what follows we provide results of the capture rates by numerically integrating the cross section according to Eq.~\ref{averagesigma}. 

\begin{figure}
\centering
 \includegraphics[width=0.8\linewidth]{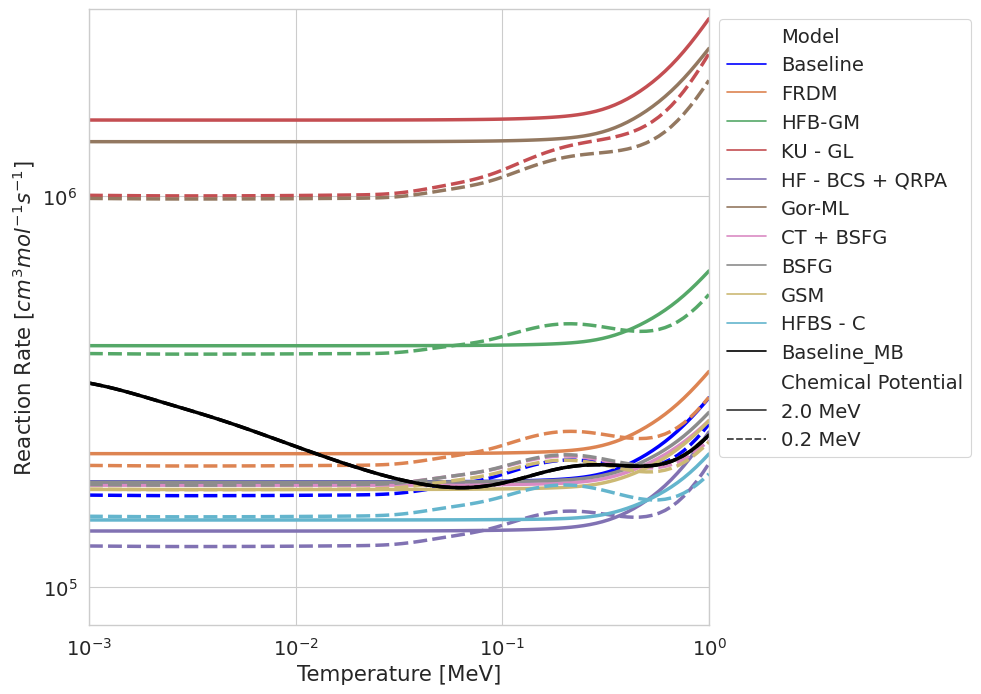}
   \caption{Reaction rates for the capture of degenerate neutrons on $^{80}$Ge as a function of temperature, when the nuclear model inputs are varied as in table 1. The chemical potential is $\mu_n$=2.0 MeV for the solid lines, and  $\mu_n$ = 0.2 MeV  for the dashed lines. %The REACLIB rate for non-degenerate neutrons based on the NON-SMOKER Hauser-Feshbach code is included for comparison. 
   }
   \label{fig:80Ge} 
\end{figure}

Our results are based on capture cross sections obtained via the nuclear reaction code TALYS 1.9 \cite{talysdocumentation,talys}, following the Hauser-Feshbach statistical treatment of the compound nucleus. We use the Hartree-Fock-Bogoliubov
with Skyrme forces mass model (HFB-SM) \cite{HFB-SM}, the Skyrme-Hartree-Fock-Bogoliubov model with  quasiparticle random-phase approximation $\gamma$-strength function (HFB+QRPA) \cite{HFB-QRPA}, and the Hartree Fock using Skyrme force nuclear level density model (HFS) \cite{HFS}.  This model parameterization will be referred to as the "baseline" input
We find the rates by integration of the TALYS cross sections for a range of temperatures and neutron chemical potentials relevant to the conditions in a neutron start crust. As an example, Figure \ref{fig:80Ge} shows the resulting capture rates on $^{80}$Ge with the baseline input, as well as other variations in model parameters (see section 2.2).

\subsection{Impact of Neutron Degeneracy}
\label{sec:distribution}
To asses the effect of neutron degeneracy we follow Shternin et al \cite{Shternin:2012pt} by calculating the ratio

\begin{equation}
  R = \frac{\langle \sigma v\rangle_{FD}}{\langle \sigma v\rangle_{MB}},
  \label{eq:ratio_simple}
\end{equation}
where $\langle \sigma v\rangle_{FD(MB)}$ is the averaged cross section from Eq.~\ref{averagesigma} over the Fermi-Dirac (Maxwell-Boltzmann) distribution function. Here we provide results of $R$ by numerically integrating the cross section accordingly with $f(E)$ going beyond the power law approximation used in  \cite{Shternin:2012pt} providing significantly more accurate rates \cite{Knight:2019UNI}. 

\begin{figure}[h]
\centering
   \includegraphics[width=0.8\linewidth]{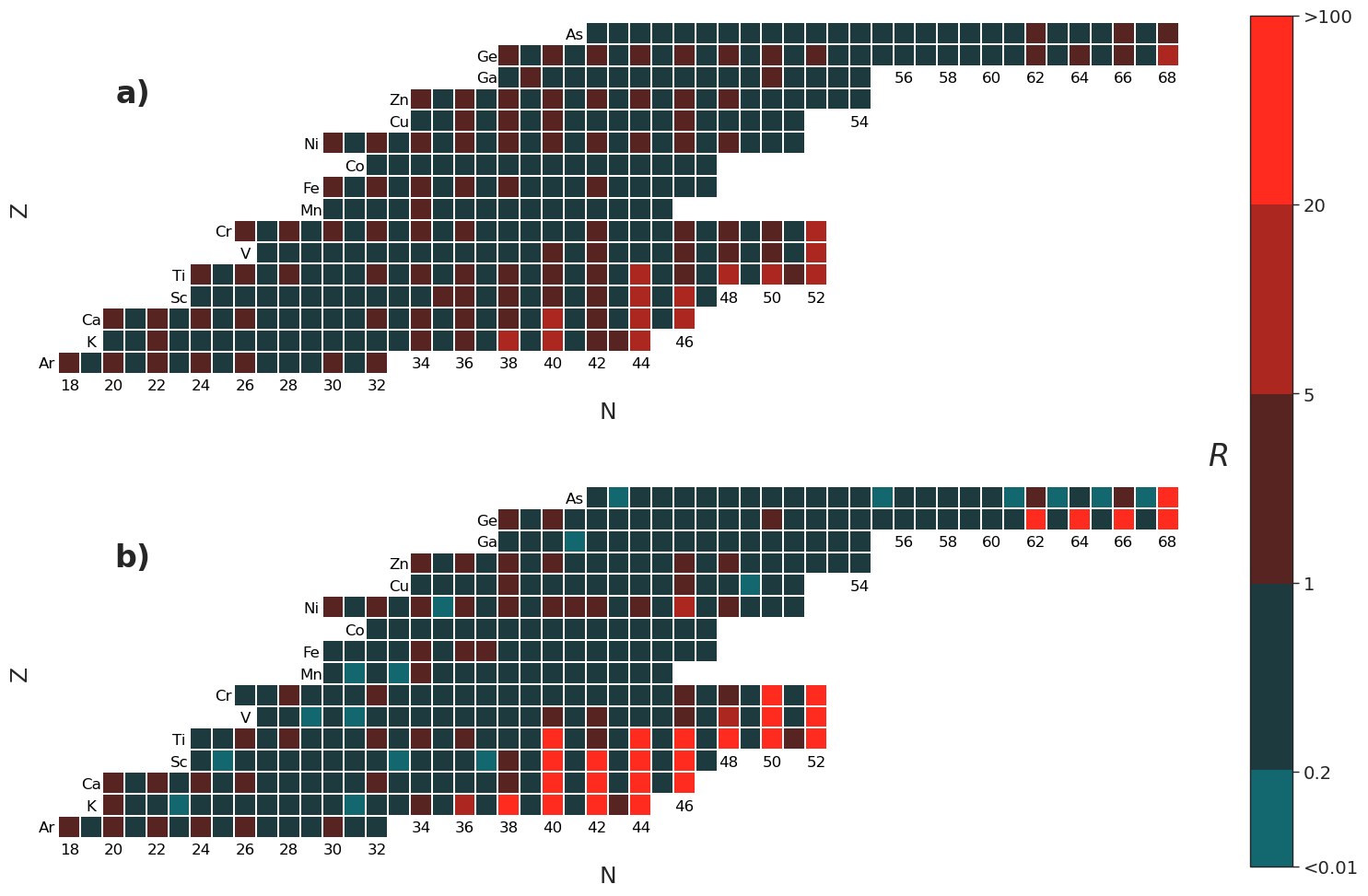}

\caption[Maxwell Rates SL]{Ratio, $R$, between degenerate and classical capture rates using the baseline input discussed in the text. (a) shows reaction rates at 1.5 GK and chemical potential 0.5 MeV (b) shows reaction rates at 0.2 GK and a chemical potential of 2.0 MeV}
\label{fig:R}
\end{figure}

Figure \ref{fig:R} shows the results for $R$ based on the baseline nuclear input parameters. Particularly large changes due to degeneracy are found for very neutron rich nuclei near the neutron drip line. As expected, the effect is even more pronounced for lower temperatures and larger neutron chemical potential. For some nuclei, e.g. $^{74}$Ti (Z = 22, N = 52) the ratio can grow to many orders of magnitude (as large as $10^{46}$). These large changes are for neutron capture rates with negative Q-values where $T < |Q|$ but $\mu_n > |Q|$. In such cases, the Maxwell-Boltzmann distribution rates are limited to the few neutrons in the high energy tail of the distribution, while large numbers of neutrons are available with the degenerate Fermi-Dirac distribution. Such large enhancements are therefore particularly seen in nuclei near the drip line with an even number of neutrons, which tend to have very low or negative neutron capture Q-values resulting in the pronounced staggering of $R$ for odd vs even neutron numbers in Fig. 2. Closer to stability away from the drip line the differences are relatively small in most cases, reflecting the 1/$v$ trend of the standard s-wave cross section that makes rates energy independent. There are however many exceptions with deviations of factors of up to 5 due to the deviations from the 1/$v$ behavior caused by resonances at higher energies. This leads typically to an enhancement of the degenerate Fermi-Dirac distribution based rates, as those have higher neutron energies and thus sample higher energy parts of the cross section. There are a few exceptions where the degeneracy leads to a decrease of the capture rates. These are cases where the neutron capture cross section drops faster than 1/$v$ at high energies, for example due to the opening of additional channels. 

\subsection{Uncertainties due to nuclear model input}
\label{sec:nuclear physics}
 Calculations of neutron capture rates is subject to uncertainties related to the  theoretical predictions of the reaction cross sections. The complex character of the nuclear force and the intractability challenge posed by the nuclear many-body problem have resulted in the development of a variety of methods and models predicting nuclear properties.  
TALYS allows the implementation of different theoretical models for the inputs needed in reaction calculations, making the comparison between such nuclear physics inputs possible.
For the results discussed in this section, we use a neutron chemical potential of 1.00 MeV and a range of temperatures between 0.001 and 1 MeV, (for reference, at a temperature of 0.001 MeV and neutron density 1.6$\times 10^{-4}$ fm$^3$ the neutron thermal wavelength is ten times larger than the average neutron free gas spacing). Further lowering the temperature does not provide any "new" information for model comparison. 

We calculated neutron capture rates with different nuclear physics input, as summarized in Table \ref{tab:model list}, and compare to the "baseline" calculation (first row in Table \ref{tab:model list}).  Starting from the baseline, one nuclear model was changed at a time, resulting in 10 cross sections as functions of energy for each nucleus. Note that the HFB-SM mass model uses the parameter set Bsk17 \cite{HFB-SM} of the Skyrme force  , and HFB-GM mass model uses the D1M parameterization of the Gogny force \cite{HFB-GM-P}.
The neutron capture rates were then calculated for each cross section with incident neutron velocities that follow a Maxwell-Boltzmann(MB) or Fermi-Dirac (FD) distribution. We analyzed the range of variations resulting from changes in level density and $\gamma$-strength function separately from the mass model variation to facilitate comparison with \cite{Liddick:2016eot}, who performed a similar study for neutron captures based on Maxwell-Boltzmann distributions. 
%We have also included the corresponding REACLIB rate \cite{Cyburt:2010} based on the NON-SMOKER Hauser-Feshbach code \cite{Rauscher:2000fx}. 

\begin{table}
\caption{List of nuclear physics models used.}
\scriptsize
\begin{tabular}{@{}lll}
%  \toprule[1.5pt]
\br
  \textbf{Nuclear Level Density} &\textbf{$\gamma$-strength function} &\textbf{ Mass models} \\ 
%  \toprule[1.5pt]
\br
Hartree Fock using Skyrme force (HFS) \cite{HFS}   & Hartree-Fock-Bogoliubov + QRPA  &   Hartree-Fock-Bogoliubov  \\
   &  (HFB + QRPA)\cite{HFB-QRPA}    & using Skyrme forces (HFB - SM) \cite{HFB-SM} \\
   \bs
   \hline
\bs
  Back-shifted Fermi gas model (BSFG) \cite{BSFG}  & Kopecky-Uhl Generalized Lorentzian  & FRDM\cite{FRDM}   \\
    &(KU-GL) \cite{Ku-Gl}  &  \\
 \bs
 Generalized super fluid model (GSM) \cite{GSM}  & Hartree-Fock BCS  + \ QRPA    &  Hartree-Fock-Bogoliubov   \\
   &  (HF-BCS + QRPA) \cite{HF-BCS-QRPA}   &   using Gogny forces(HFB - GM) \cite{HFB-GM, HFB-GM-P} \\
 \bs
Constant temperature matched to  & Modified Lorentzian (Gor-ML) \cite{Gor-ML}   &   \\
 the Fermi gas model (CT + BSFG)\cite{CT-BSFG} &   &    \\
 \bs
  Hartree-Fock-Bogoliubov (Skyrme force)  &   &   \\
   + combinatorial method (HFBS - C) \cite{HFBS-C}  &   &   \\
 \br

\end{tabular}

\label{tab:model list}
\end{table}
\normalsize

\begin{figure}
\centering
   \includegraphics[width=0.8\linewidth]{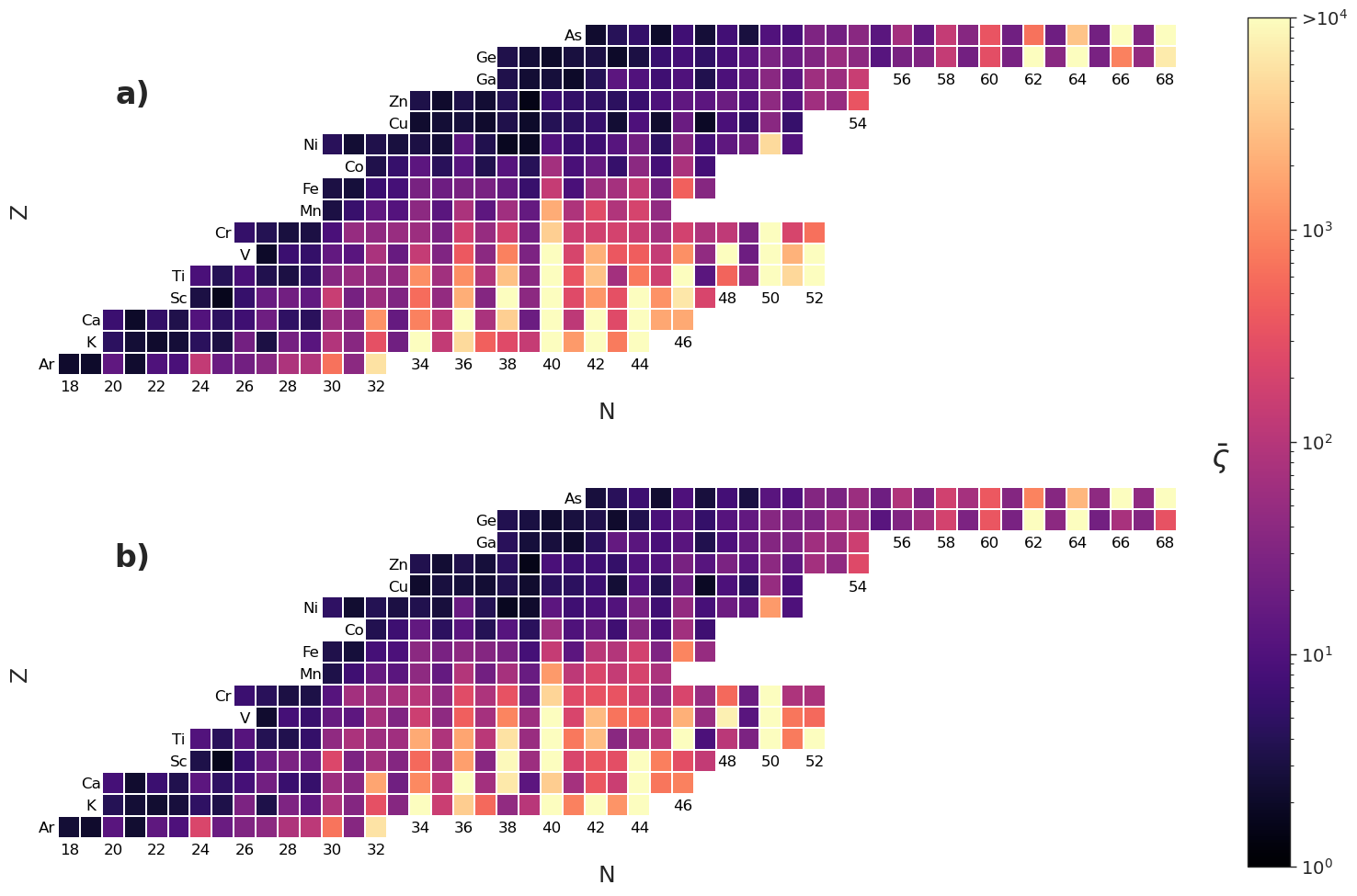}

\caption[Fermi Rates MSL]{$\bar{\varsigma}$ for degenerate neutron capture rates with variations in nuclear mass, nuclear density level, and gamma strength function models. Figure (a) shows reaction rates at 1.5 GK and a chemical potential of 0.5 MeV (b) shows a similar case, but for a colder, denser system at 0.2 GK and a chemical potential 2.0 MeV.}
\label{SD_MDS}
\end{figure}

\begin{figure}
\centering

   \includegraphics[width=0.8\linewidth]{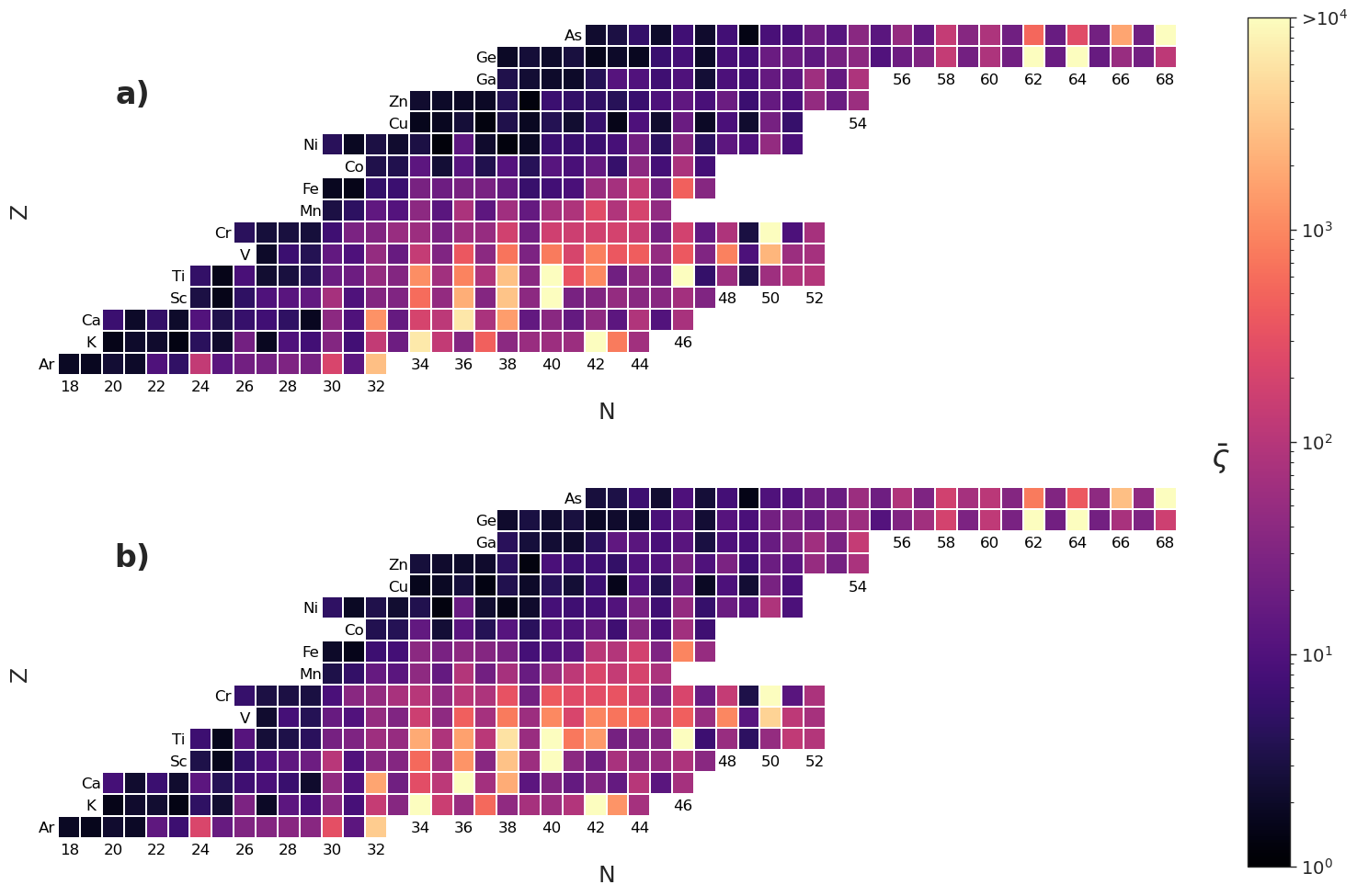}

\caption[Fermi Rates SL]{$\bar{\varsigma}$ for degenerate neutron capture rates with variations in nuclear density level and gamma strength function models. Figure (a) shows reaction rates at 1.5 GK and a chemical potential of 0.5 MeV (b) shows a similar case, but for a colder, denser system at 0.2 GK and a chemical potential 2.0 MeV.}
\label{SD_DS}
\end{figure}

\begin{figure}
\centering

   \includegraphics[width=0.8\linewidth]{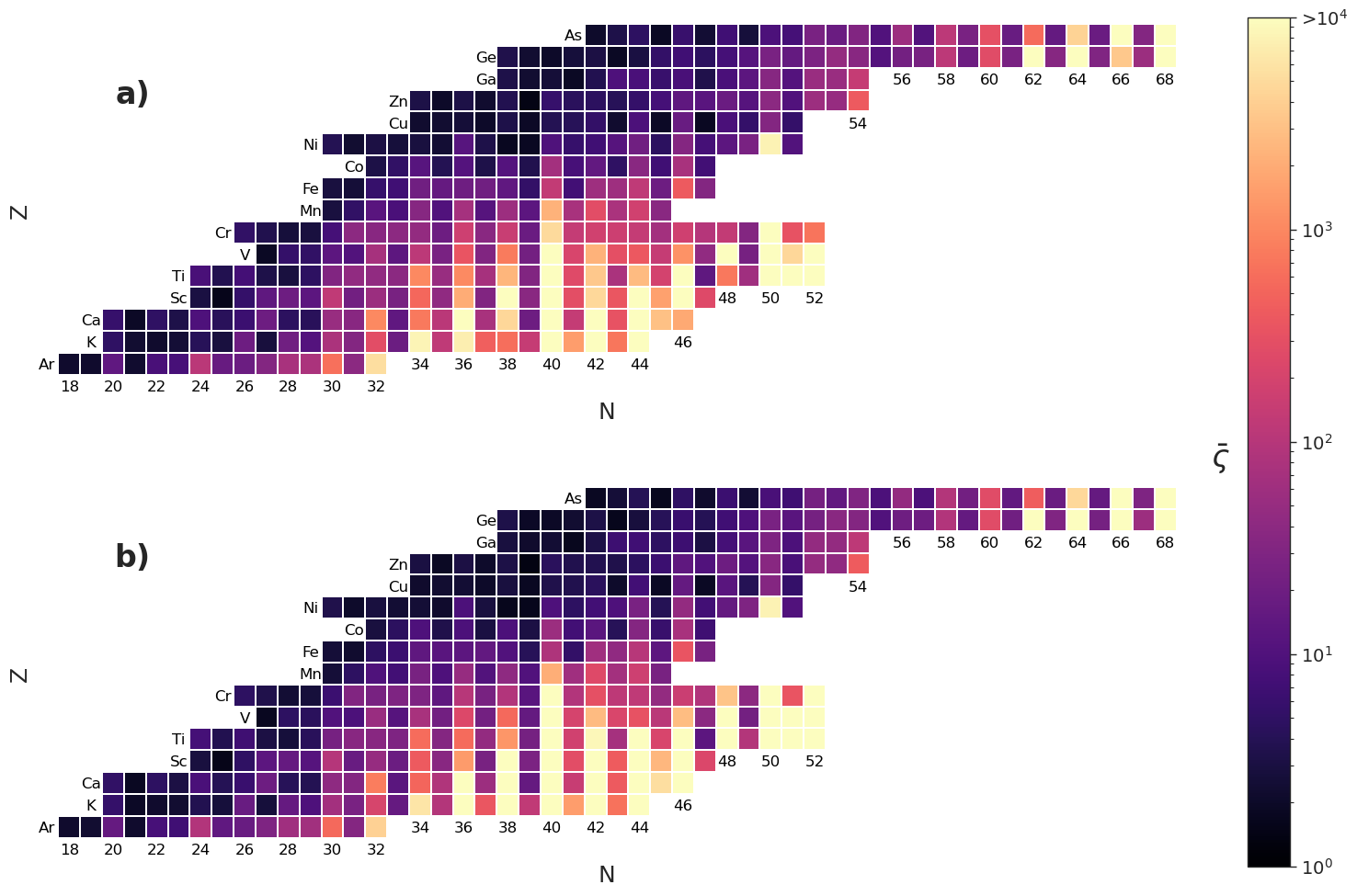}

\caption[Maxwell Rates MSL]{$\bar{\varsigma}$ for classical thermonuclear neutron capture rates with variations in nuclear mass, nuclear level density, and gamma strength function models. (a) shows reaction rates at 1.5 GK while (b) shows reaction rates at 0.2 GK}
\label{fig:MB-SD_MDS}
\end{figure}

\begin{figure}
\centering

   \includegraphics[width=0.8\linewidth]{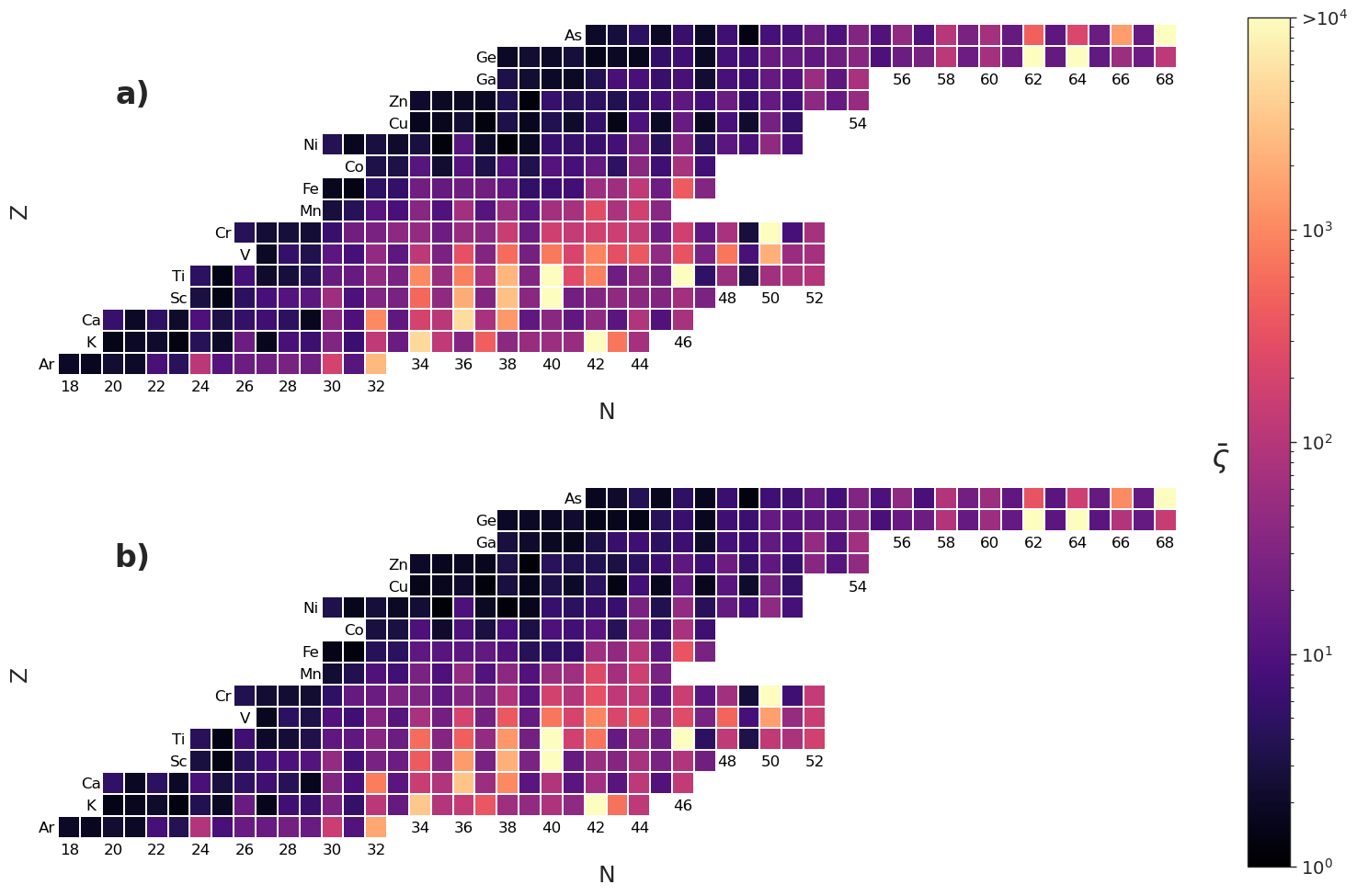}

\caption[Maxwell Rates SL]{$\bar{\varsigma}$ for classical thermonuclear neutron capture rates with variations in nuclear level density and gamma strength function models. (a) shows reaction rates at 1.5 GK while (b) shows reaction rates at 0.2 GK}
\label{fig:MB-SD_DS}
\end{figure}

To quantify the sensitivity of each $(n,\gamma)$ rate to nuclear model variations, we chose to use the ratio between the largest and smallest rate within the set of chosen nuclear models: $\bar{\varsigma}= \max/\min$. Figure \ref{SD_MDS} shows the $\bar{\varsigma}$ values for degenerate reaction rates using a set in which all three nuclear properties where changed, while Figure \ref{SD_DS} shows results when only the nuclear level density and $\gamma$-strength function were varied. The top and bottom panels have different values of temperature $T$ and neutron chemical potential $\mu_n$, with the conditions of the bottom panel corresponding to higher degeneracy.  Figures \ref{fig:MB-SD_MDS} and \ref{fig:MB-SD_DS} show for comparison $\bar{\varsigma}$ values obtained with a classical Maxwell-Boltzmann distribution.

%-------------
%-------------

%-------------
%-------------

%-------------
%-------------

\newpage
%-------------------------------------------------------
Similar to what has been found for neutron capture rates with non-degenerate neutrons, input model induced uncertainties in degenerate neutron capture rates grow dramatically with distance from stability, reaching in many cases 3-4 orders of magnitude. When mass model variations are included, the number of cases with very large dispersion increases significantly due to the changes in reaction Q-values. Near the neutron drip line, a reaction can be endothermic according to one mass model but not another. If the reaction is endothermic (i.e. a negative Q value), there is a threshold energy required to start the reaction. Thus, if neutrons do not have the requisite thermal energy, the reaction rate is very close to zero. This gives rise to the very large variations in reaction rate predictions. Captures on even $N$ nuclei are more sensitive to variations than those on odd $N$ nuclei, resulting in a staggering effect. When the mass model is varied, reactions with small Q-values result in large relative changes between models.  For reactions with Q-values around zero, a variation can flip the reaction from exothermic to endothermic, making the change even larger.  In contrast, mass model variations cause smaller relative changes in larger Q-values, and thus smaller rate variations. When the mass model is kept fixed, and only variations in the density level and gamma strength functions are performed, staggering is smaller but still pronounced. The largest variations are due to
significant differences
between the predictions from phenomenological and microscopic models of the gamma strength function around neutron separation energy. 

\begin{figure}
    \centering
    \includegraphics[width=0.8\textwidth]{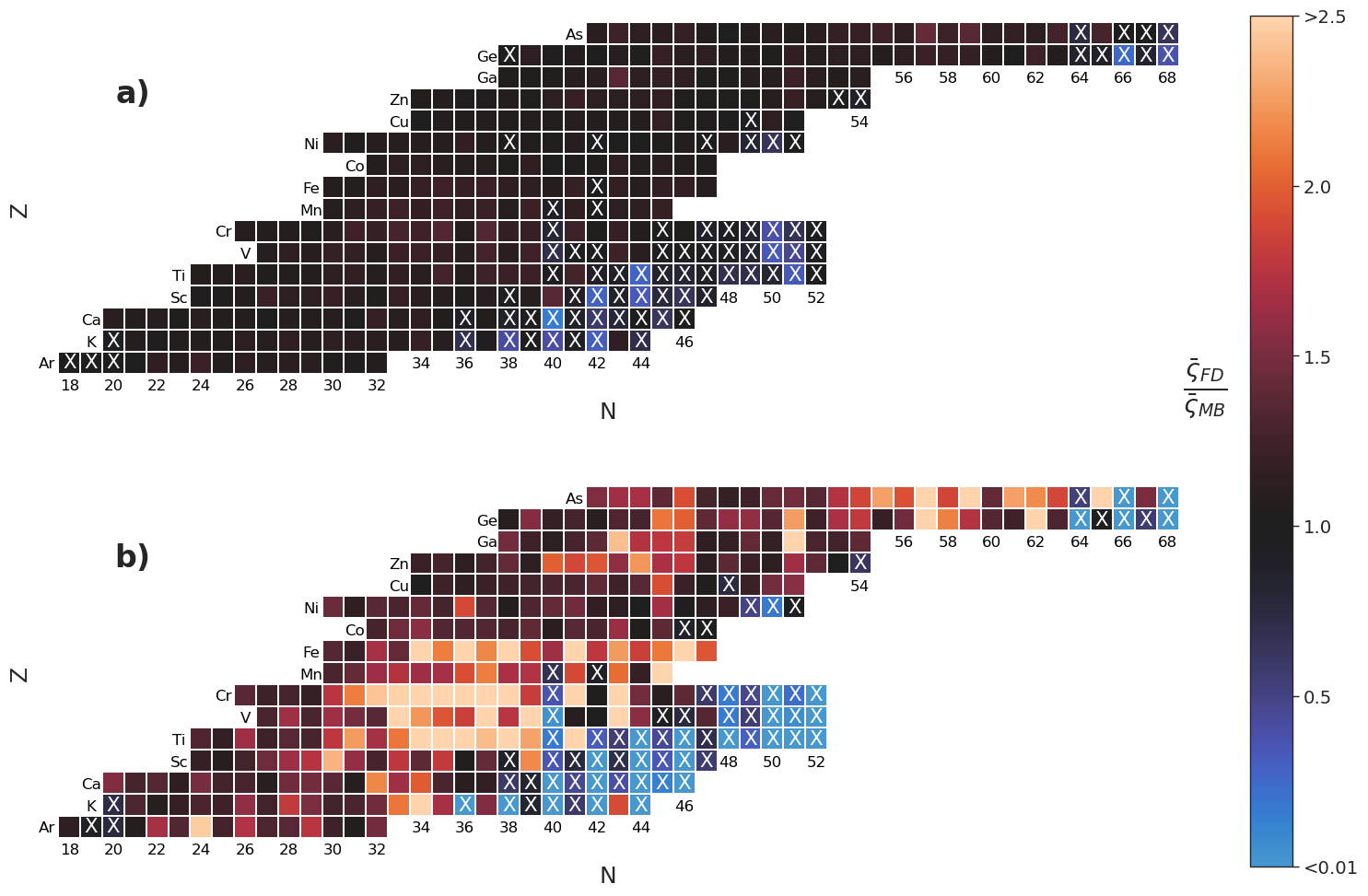}

    \caption[Differences-MSL]{$\frac{\bar{\varsigma}_{FD}}{\bar{\varsigma}_{MB}}$ calculated at a temperature of 1.5 GK and chemical potential of 0.5 MeV (upper panel), and at $T=$0.2 GK and chemical potential of 2.0 MeV (bottom panel). An X denotes cases where $\bar{\varsigma}_{MB} > \bar{\varsigma}_{FD}$.}
    \label{fig:diff-MSL}
\end{figure}

\begin{figure}
    \centering
    \includegraphics[width=0.8\textwidth]{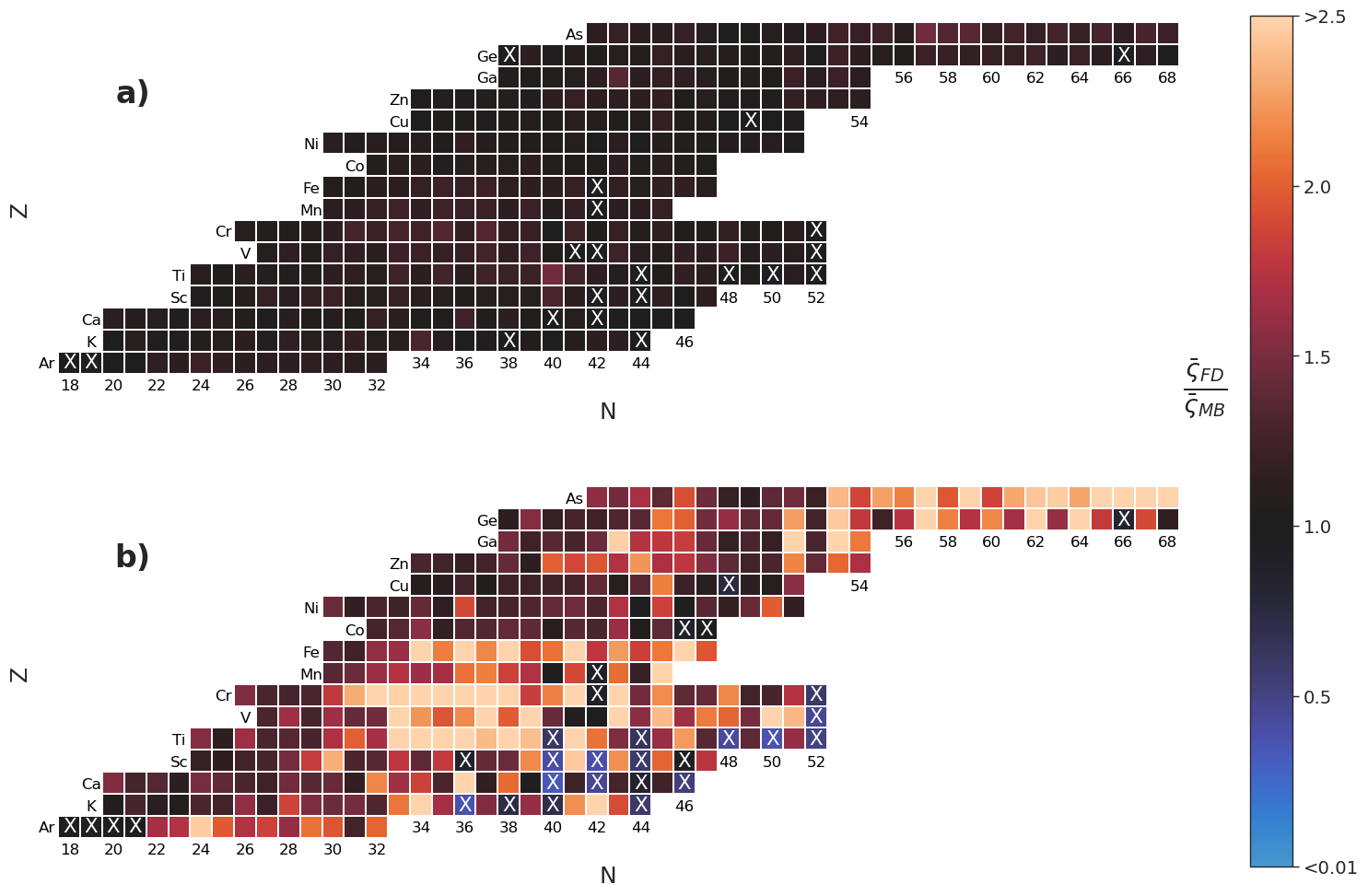}

    \caption[Differences-MSD]{Same as in Figure \ref{fig:diff-MSL} but excluding variations in the mass model.}
    \label{fig:diff-SL}
\end{figure}

 Figures \ref{fig:diff-MSL} and \ref{fig:diff-SL} illustrate the dependence of the input model induced uncertainties $\bar{\varsigma}$ on degeneracy with, and without, variations in nuclear masses, respectively. Typically degeneracy increases the dispersion of rates due to model variations significantly, especially for very degenerate conditions where changes in sensitivity can reach factors of 5 or more. The reason for the enhanced sensitivity is the larger role of higher energy neutrons that sample higher energy parts of the cross section that are more sensitive to nuclear model inputs compared to the simple $1/v$ behavior at low energy. Near the neutron drip line the situation is different, and degeneracy reduces the sensitivity to input models, especially when mass model variations are included. In these cases low energy (lower than 0.05 MeV) cross section differences are dampened by the Fermi-Dirac distribution, which samples the cross section up to 2~MeV, but are further enhanced by the Maxwell-Boltzmann distribution that favors lower energies. Near the drip line, variations in mass models can also dramatically change the threshold energy of a reaction, a change that classical reactions are particularly sensitive to. 

%===========================================================================================

\newpage
%%%%%%%%%%%%%%%%%%%%%%%%%%%%%%%%%%%%%%%%%%%%%

\section{Intermediate-energy cross section fit}
\label{fit}

Shternin et al \cite{Shternin:2012pt} proposed the use of a power law approximation for the neutron capture cross section
\begin{equation}
    \sigma(E) = \sigma_a (E-E_0)^\nu,
\end{equation}
at low energies, which would be valid up until a maximum energy $E_{max}$. In this approximation $\nu$, $E_0$ and $\sigma_a$ are used as input parameters. Note, that the ratio $R$, discussed in the previous section, is independent of $\sigma_a$. This solution provides a computationally efficient way to determine degenerate reaction rates. With the introduction of dimensionless parameters $x_0 = E_0 /T$ and $y =\mu_n /T$, Shternin et al arrived at an analytic ratio between degenerate and classical neutron capture rates: 
\begin{equation}
R=\frac{\exp x_{0}}{x_{0}+\nu+1} \frac{(\nu+1) \mathcal{F}_{\nu+1}\left(y-x_{0}\right)+x_{0} \mathcal{F}_{\nu}\left(y-x_{0}\right)}{\mathcal{F}_{1 / 2}(y)}
\label{eq:R}
\end{equation}
where $\mathcal{F}_{\nu}\left(y\right)$ is the Fermi-Dirac integral 
\begin{equation}
    \mathcal{F}_{\nu}(y)=\frac{1}{\Gamma(\nu+1)} \int_{0}^{\infty} \frac{x^{\nu} \mathrm{d} x}{1+\exp (x-y)}
\end{equation} 
and $\Gamma(\nu + 1)$ is the Euler-gamma function.
Defined this way, $R$ allows a degeneracy correction for already established rates at high temperatures. 

According to the authors the approximation is valid as long as the sum of the temperature $T$ and the maximum value among $E_0$ and $\mu_n$ is lower than $E_{max}$. However, as discussed in \cite{Knight:2019UNI}, the determination of $E_{max}$ is ambiguous and slight changes in $E_{max}$ can lead to different power indexes $\nu$. Furthermore, $E_{max}$ values can be fairly low for exothermic reactions, and therefore the power law description fails to capitalize on the compound nucleus contributions to the cross section, a region of energies accessible to neutrons, given their chemical potential in dense matter.

To alleviate the issues mentioned above, we propose an additional term to the cross section for exothermic reactions ($E_0 = 0$) to permit the use of a fit for higher chemical potentials to TALYS cross sections. We use the same power law approximation for low energies, but include $N$ Lorentzian terms to describe compound nucleus contributions above slow neutron energies:

\begin{equation}
\label{Xsection_fit}
\sigma(E)=\sigma_{a} E^{\nu} +  \sum_{i=1}^{N}\frac{\sigma_{i}}{\pi} \frac{ \tfrac{1}{2}\Omega_i}{\left(E-E_{i}\right)^{2} + {(\tfrac{1}{2}\Omega_{i})}^2}
\end{equation}
where  $\sigma_i$, $\Omega_i$ and $E_i$ are additional input parameters. We emphasize again that following Shternin et al. \cite{Shternin:2012pt}, we neglect here contributions from captures on excited target states as discussed in section 2. We have found sufficient coverage using $N = 2$ peaks, and we provide coefficients under this assumption. It should be noted the number of peaks is adaptable to the cross section data resolution. 

For the case of endothermic reactions ($E_0 > 0$), the power law fit alone is sufficient to describe reaction rates at $\sim$2 MeV chemical potential values. Essentially, without the low energy spectrum, a power law fit has enough flexibility to approximate the cross section, as long as the chemical potential is less than the energy associated with the cross sections maxima. 

\begin{figure}
\label{fig:Cross_sections}
\centering
\begin{subfigure}[b]{0.48\textwidth}
    \centering
   \includegraphics[width=\textwidth]{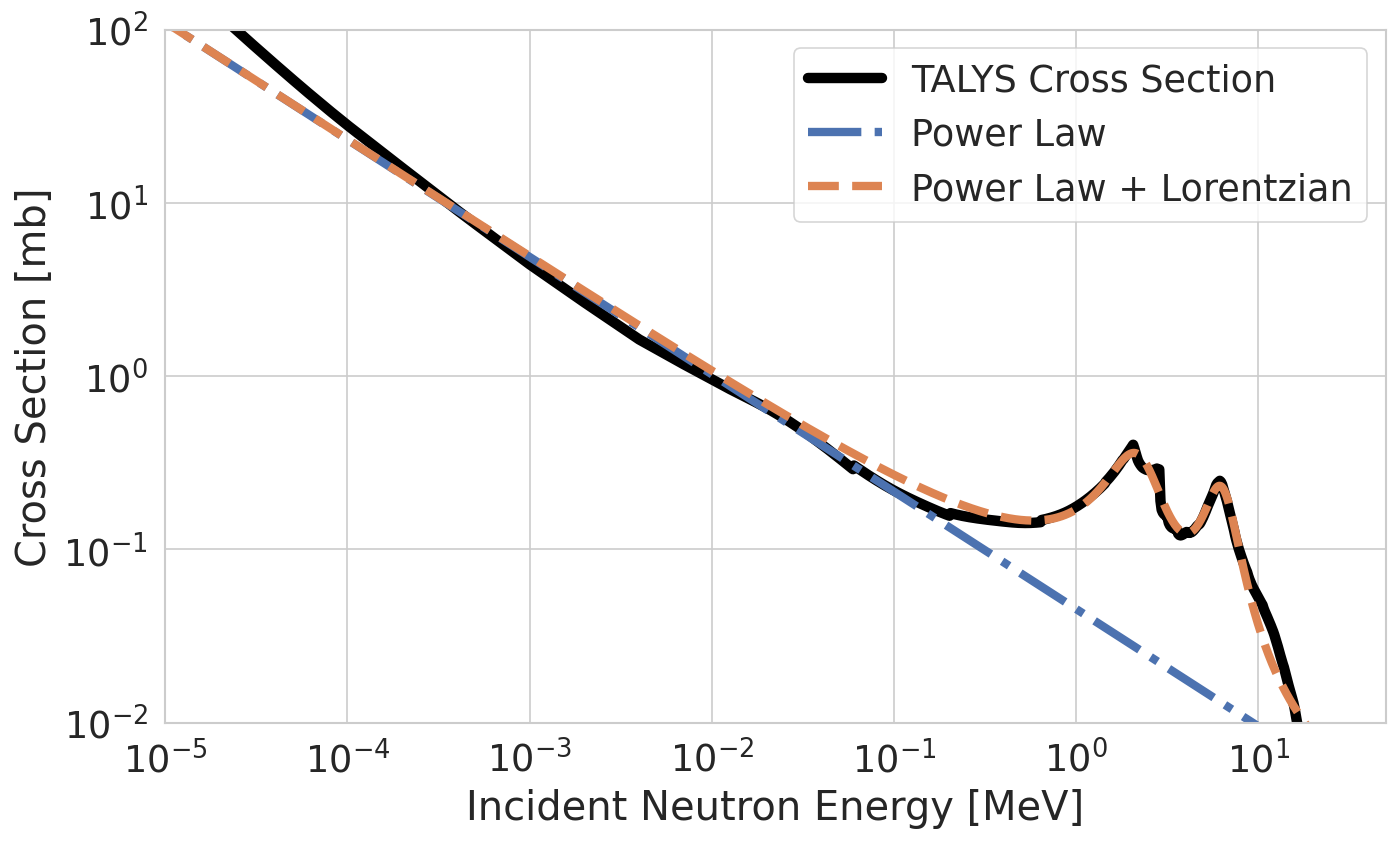}
   \caption{}
   \label{fig:Ng1} 
\end{subfigure}
\hfill
\begin{subfigure}[b]{0.48\textwidth}
\centering
   \includegraphics[width=\textwidth]{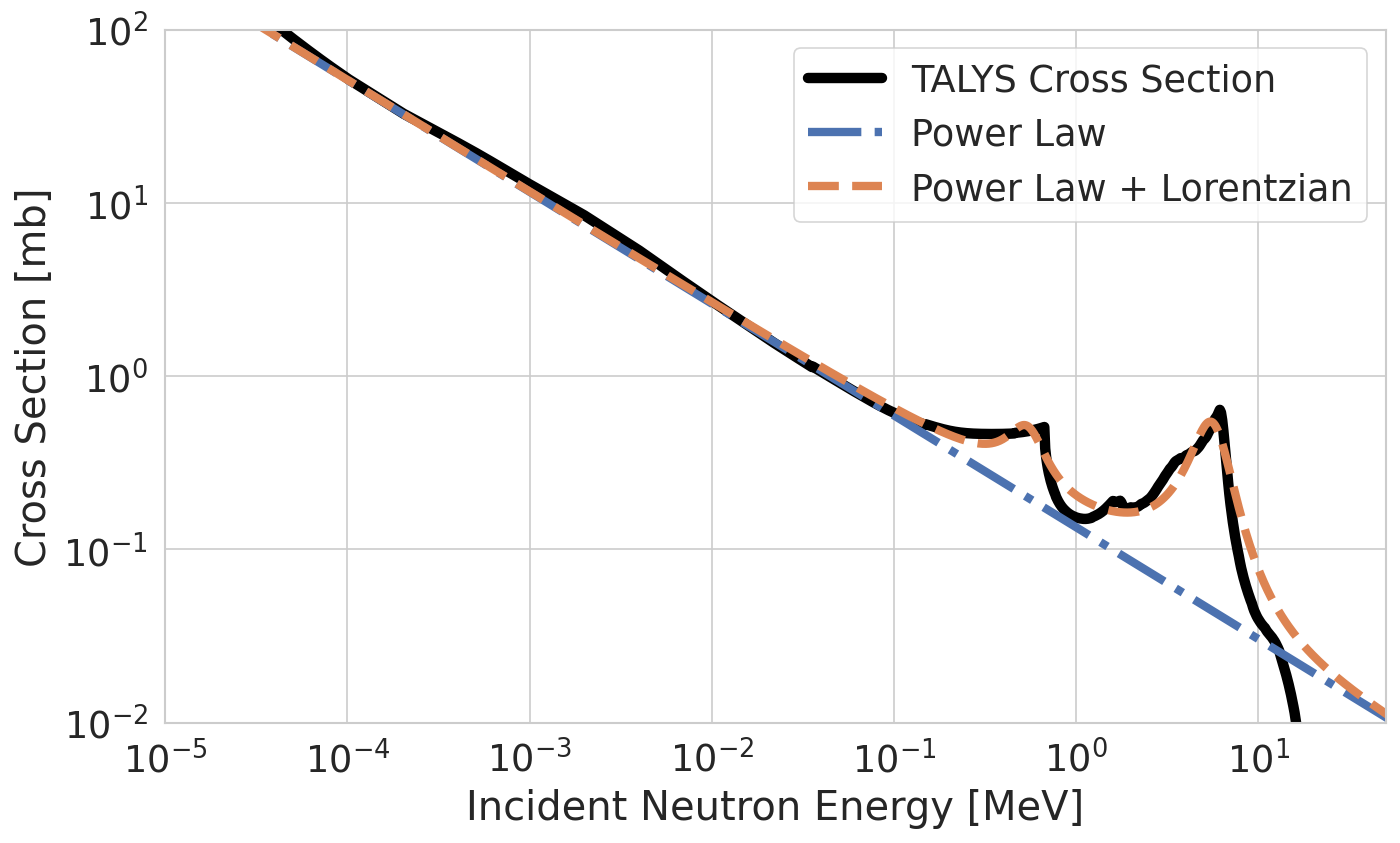}
   \caption{}
   \label{fig:Ng2}
\end{subfigure}
\caption{TALYS calculated cross-sections and corresponding fits for (a) $^{68}$Ni and (b) $^{80}$Ge , a reasonable approximation first proposed in Ref. [19]} 
\label{fig:Xfits}
\end{figure}

Figure \ref{fig:Xfits} shows a sample fit using our approximation for $^{68}$Ni and $^{80}$Ge. To determine the parameters in Eq.~\ref{Xsection_fit}, we performed weighted non-linear least squares. The weighting scheme is motivated by the integrand in Eq.~\ref{averagesigma}. The integrand itself is directly proportional to the incident neutron energy. Thus, our weighting coefficients are proportional to incident neutron energy, which will put a bias towards fitting peaks at higher energies. However, to ensure the power law behaviour has adequate attention, we restrict our fit to only use a power law until an incident energy of 0.01 MeV. After this point, the power law coefficients are fixed, and the Lorentzian terms are then optimized over the entire domain of incident energies.

From Eq.~\ref{Xsection_fit} and Eq.~\ref{averagesigma}, we can use the Sommerfeld approximation to obtain degenerate reaction rates in the low temperature limit. We justify the use of a Sommerfeld approximation as the additional Lorentzian terms are included to explicitly account for larger chemical potential values.  Defining  additional dimensionless parameters $\omega_i = \tfrac{\Omega_i}{2T}$ and $x_i = \tfrac{E_i}{T}$, we obtain

\begin{equation} 
   \begin{split}
   \langle \sigma v\rangle_{FD}= &
   \sqrt{\frac{8}{\pi m T}} \frac{1}{\mathcal{F}_{1 / 2}(y)} 
   \Bigg[ \sigma_a T^{\nu + 1}\Gamma(\nu + 2) \mathcal{F}_{\nu + 1}(y) +
   \sum_{i=1}^{N} 
   \frac{\sigma_{i}}{\pi} \Bigg[ \frac{\omega_{i}}{2} \ln \left(\frac{\left(x_{i}-y\right)^{2}+\omega_{i}^{2}}{x_{i}^{2}+\omega_{i}^{2}}\right)+ \\
   & x_{i} \left( \arctan \left(\frac{x_{i}}{\omega_{i}}\right)- \arctan \left(\frac{x_{i}-y}{\omega_{i}}\right) \right) + \frac{\pi^2}{6} \frac{\omega_{i}\left(\omega_{i}^{2}-y^{2}+x_{i}^{2}\right)}{\left(\omega_{i}^{2}+\left(y-x_{i}\right)^{2}\right)^{2}} \Bigg] \Bigg] .
   \end{split}
   \label{eq:analytic}
\end{equation}

To quantify our error, we use a normalized root-mean-square error (NRMSE) relative to numerically calculated degenerate reaction rates, based directly on TALYS cross sections, at fixed temperatures and between chemical potentials of 0.2 and 2.0 MeV. The normalization is with respect to the mean of the numerical reaction rate within the range of chemical potential values per nuclei. 

\begin{figure}
    \centering
    \includegraphics[width=0.9\textwidth]{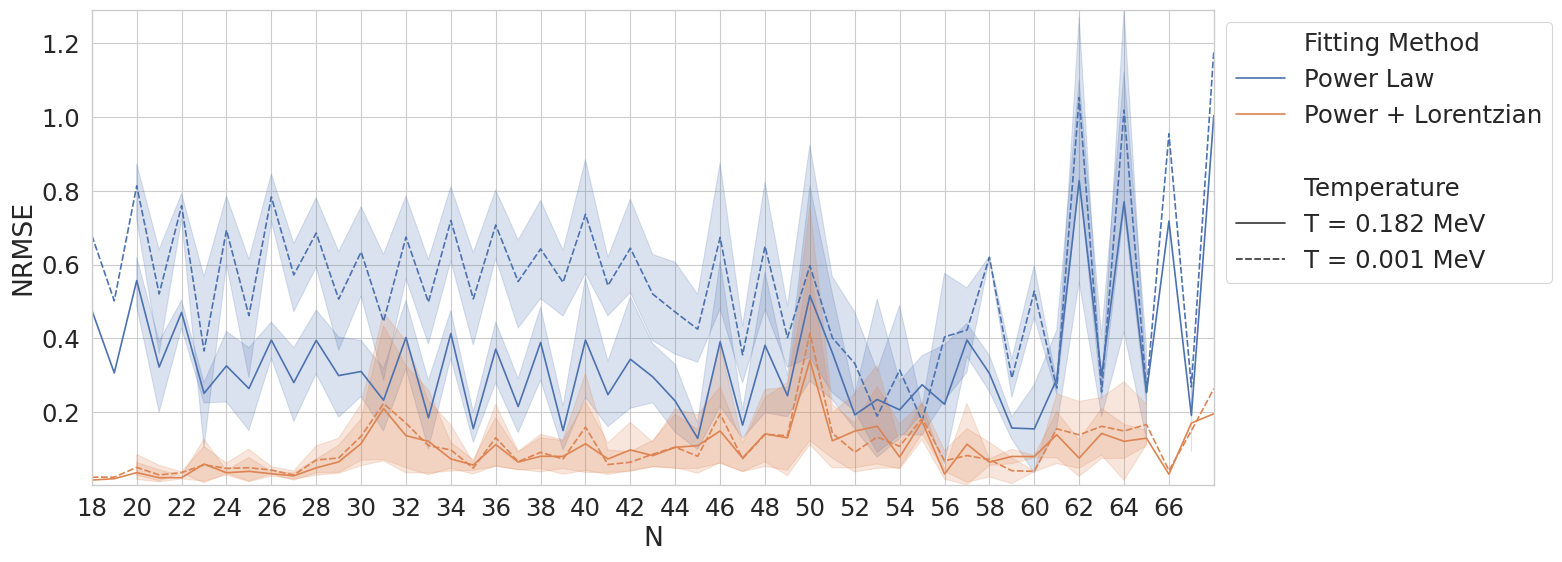}
    \caption[NRMSE]{NRSME values averaged over neutron number for degenerate neutron captures}
    \label{fig:NRMSE_aggregated}
\end{figure}

Figure \ref{fig:NRMSE_aggregated} shows NRMSE values for exothermic reactions aggregated over neutron number for the two approaches : power law (only) fit and power law plus Lorentizan peaks fit (Eq. \ref{eq:analytic}).The shaded areas correspond to the error of the NRMSE. Note that larger fluctuations at higher $N$ is primarily due to less counts of neutron rich nuclei in the dataset.
We see that not only does a Lorentzian addition improve the relative error by about 30\%, but it also has a significantly better temperature-dependent performance. 

\newpage

\begin{figure}
    \centering
    \includegraphics[width=0.9\textwidth]{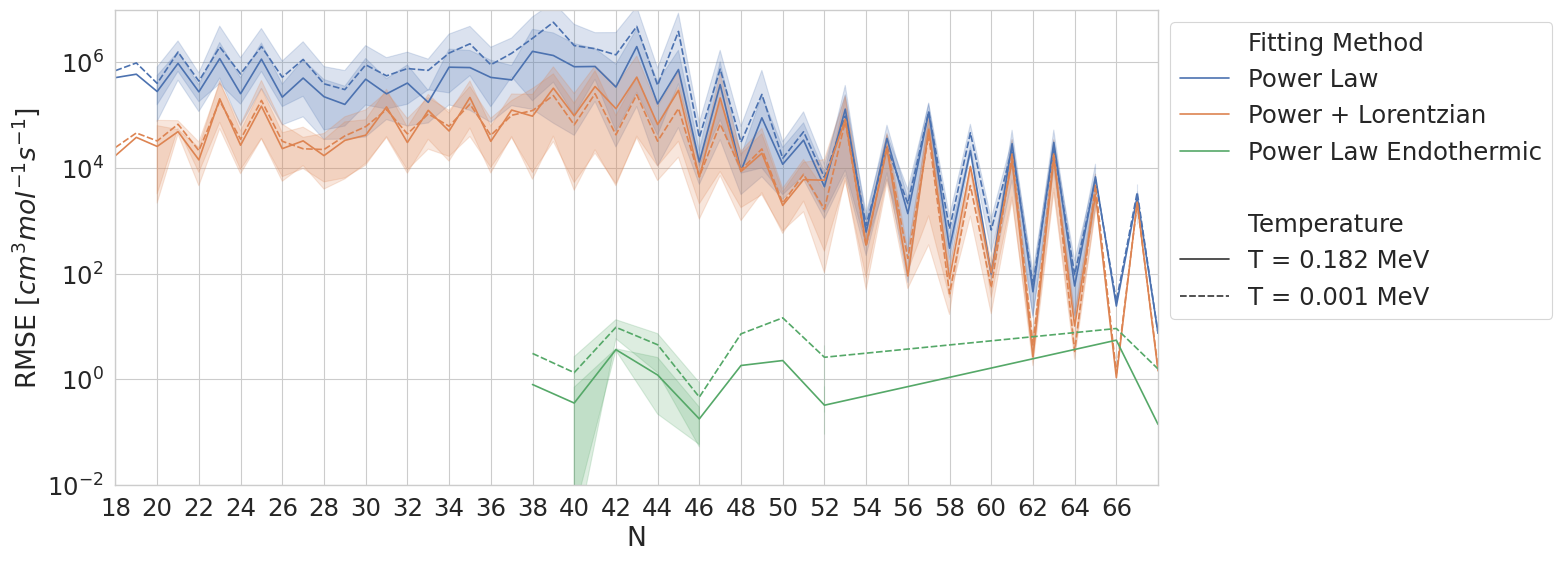}
    \caption[RMSE]{ Absolute RSME values averaged over neutron number for degenerate neutron captures}
    \label{fig:RMSE_aggregated}
\end{figure}

The absolute (non-normalized) RMSE values are shown in Figure \ref{fig:RMSE_aggregated}. The endothermic reaction rates have much smaller absolute errors, and a power law approach is sufficient for those reactions.

%%%%%%%%%%%%%%%%%%%%%%%%%%%

%%%%%%%%%%%%%%%%%%%%%%%%%%%

%%%%%%%%%%%%%%%%%%%%%%%%%%%

\section{Discussion}
\label{sec:discussion}

We have presented results of the effects of degeneracy on neutron capture rates on neutron rich nuclei and on uncertainties of Hauser-Feshbach model predictions of these rates as determined by variations from different choices of the nuclear input models. Our results are in qualitative agreement with previous studies of non-degenerate neutron capture rates \cite{Liddick:2016eot} who found dramatic increases in $\gamma$-strength and level density input model  induced variations in Hauser-Feshbach calculations away from stability, increasing from factors of 5-10 a few mass units away from stability to $>$100 at around 10 mass units. We show that these uncertainties grow to 3 to 4 orders of magnitude as one approaches the neutron drip line and that they are further enhanced for capture of degenerate neutrons.  Note that Beard et al \cite{Beard:2014jaa} found that different code implementations can result in changes of 20\%  in the Maxwell averaged cross section. We expect that the degenerate neutron capture rates would present similar trends regarding different implementations.  
 
We have also presented a new fit for the low energy TALYS cross section, and an analytic degeneracy correction to capture rates that complements the work of Shternin et al by including low energy resonances. We find that our proposed cross section paramaterization dramatically improves the accuracy for exothermic reactions compared to using a power law only. To ease the implementation degenerate neutron capture rates in nucleosynthesis studies, we provide in \cite{bryn-github}, %the supplement material 
both the fit parameters for the cross section from which rates can readily be calculated using Eq.~\ref{eq:analytic} as well as tables of the numerically integrated rates for nuclei with a range of proton numbers $3\le Z \le 85$. 

For implementations where a correction factor $R$ as calculated in \cite{Shternin:2012pt}  is used to correct a library of Maxwell-Boltzmann rates (e.g. REACLIB), a similar correction factor $R$ can be obtained if the temperature is lower than the onset of the Lorentzian terms (i.e., $T < E_i - \Omega_i$). In such cases, it is reasonable to assume the classical Maxwell-Boltzmann reaction rate $ \langle \sigma v\rangle_{MB} $ is well approximated using a power-law cross section.  $ \langle \sigma v\rangle_{MB} $ can then be calculated from the  power law fit coefficients ($\sigma_a$ and $\nu$) provided as
\begin{equation}
    \langle \sigma v\rangle_{MB} \approx 2\sqrt{\frac{2}{\pi m}}\sigma_aT^{\nu + 1/2}\Gamma(\nu + 2) \quad for \quad T < E_i - \Omega_i.
\end{equation}
With $\langle \sigma v\rangle_{FD}$ from Eq.~\ref{eq:analytic} or from the provided tables, $R$ can readily be determined. If the temperature of the system goes above this limit, we recommend to determine $R$ by numerical integration, or, alternatively, directly use the Fermi-Dirac distribution based reaction rates provided here. 

Re-evaluating the impact of neutron degeneracy on nucleosynthetic yields is beyond the scope of this work. However, our results motivate further studies about the abundances' evolution of rp-process ashes and energy generation in accreted neutron star crusts.

\ack O. L. C. acknowledges support from the National Sciences and Engineering Research Council of Canada (NSERC), and the Canada Foundation for Innovation (CFI) under the John R. Evans Leader Fund (JELF). H.S. acknowledges support from the National Science Foundation under PHY-2209429 and OISE-1927130 (IReNA).

%======================================================

\end{document}